\documentclass[graybox]{svmult}

\usepackage{mathptmx}       % selects Times Roman as basic font
\usepackage{helvet}         % selects Helvetica as sans-serif font
\usepackage{courier}        % selects Courier as typewriter font
\usepackage{type1cm}        % activate if the above 3 fonts are
\usepackage{makeidx}         % allows index generation
\usepackage{graphicx}        % standard LaTeX graphics tool
\usepackage{multicol}        % used for the two-column index
\usepackage[bottom]{footmisc}% places footnotes at page bottom

\makeindex             % used for the subject index
\begin{document}

\title*{Three dimensional structure of penumbral filaments from Hinode observations}
\author{K. G. Puschmann, B. Ruiz Cobo, V. Mart\'\i nez Pillet}
\institute{K. G. Puschmann \at Instituto de Astrof\'\i sica de Canarias,
              38200 La Laguna, Tenerife, Spain, \email{kgp@iac.es}
\and B. Ruiz Cobo \at Instituto de Astrof\'\i sica de Canarias,
              38200 La Laguna, Tenerife, Spain \email{brc@iac.es}
\and V. Mart\'\i nez Pillet \at Instituto de Astrof\'\i sica de Canarias,
              38200 La Laguna, Tenerife, Spain \email{vmp@iac.es}}

\maketitle

\vspace{-2cm}
\abstract{We analyse spectropolarimetric observations of the penumbra of the NOAA AR 10953 at high spatial resolution (0.3$^{\prime\prime}$). The full Stokes profiles of the Fe~{\sc i} lines at 630.1\,nm and 630.2\,nm have been obtained with the Solar Optical Telescope (SOT) on board the Hinode satellite. The data have been inverted by means of the SIR code, deriving the stratifications of temperature, line of sight velocity, and the components of the magnetic field vector in optical depth. In order to evaluate the gas pressure and to obtain an adequate geometrical height scale, the motion equation has been integrated for each pixel taking into account the terms of the Lorentz force. To establish the boundary condition, a genetic algorithm has been applied. The final resulting magnetic field has a divergence compatible with 0 inside its uncertainties. First analyses of the correlation of the Wilson depression with velocity, temperature, magnetic field strength, and field inclination strongly support the uncombed penumbral model proposed by Solanki \& Montavon \cite{solankimontafon93}.}

\section{Introduction}
\label{sec:1}
The filamentary structure of sunspot penumbrae still harbours many open questions: e.g. the source of the Evershed flow, the origin of dark cored penumbral filaments discovered by Scharmer et al. \cite{scharmeretal02}, as well as the structure of penumbral filaments, which could be elevated magnetic features (uncombed penumbral model by Solanki \& Montavon \cite{solankimontafon93}, Borrero et al. \cite{borreroetal05}, \cite{borreroetal06}), convective penetrations (gappy penumbral model by Spruit \& Scharmer \cite{scharmerspruit06}), or a combination of both scenarios, as suggested by MHD simulations by Rempel et al. \cite{rempeletal08}. Sophisticated inversion methods give information about the stratification of physical quantities like temperature $T$, magnetic field strength $B$, field inclination $\gamma$, field azimuth $\phi$, and line of sight velocity $V_{\rm los}$ versus optical depth, but do not provide information about geometrical heights (the Wilson depression can not be directly obtained by the inversions). Carroll \& Kopf \cite{carrolkopf08} obtain stratifications of physical quantities in geometrical height applying a neuronal network inversion technique based on MHD simulations of quiet sun. However, only information included in the MHD simulation can be retrieved. The establishment of a geometric height scale is important for the determination of the electrical current vector $\vec{J}$ (crucial for the determination of ohmic energy dissipation and important for the extrapolation of the magnetic field vector from the photosphere to the chromosphere), for the  modelling of magnetic features in 3 dimensions, and for a demonstration of the reliability of MHD simulations.
%__________________________________________________________________

\section{Observations}

\begin{figure}[h]
  \sidecaption
   %\centering
   \includegraphics[width=7.5cm]{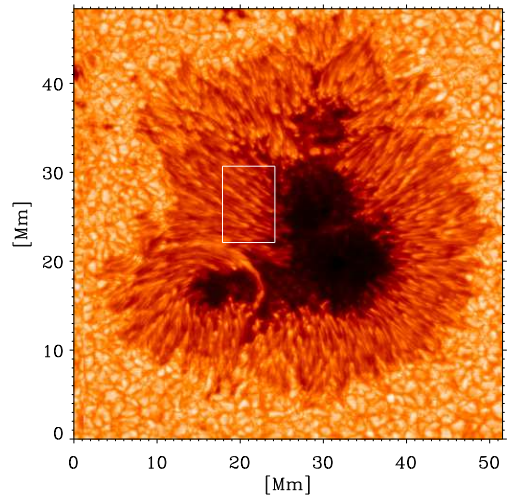}
      \caption{Continuum intensity at 630.2\,nm of a fragment of the active region AR 10953 observed by SOT on board the Hinode spacecraft. The white rectangle marks the area under study in the present paper.}
         \label{Fig1}
  \end{figure}

The active region AR 10953 was mapped at an heliocentric angle $\theta$\,=\,10$^{o}$ using the spectropolarimeter of the Solar Optical Telescope on board  the Hinode spacecraft (Lites et al. \cite{litesetal01}; Kosugi et al. \cite{kisugietal07}) on 1$^{st}$ of May 2007, between 10:46\,am and 12:25\,am UT. The region  was scanned in a thousand of steps, with a step width of 0.148$^{\prime\prime}$ and a slit width of 0.158$^{\prime\prime}$, recording the full Stokes vector of the pair of the neutral iron lines at 630\,nm with a spectral sampling of 21.53\,m\AA. The spatial resolution was $\sim$\,0.32$^{\prime\prime}$. The integration time was 4.8\,s, resulting in an approximate noise level of 1.2\,$\times$\,10$^{-3}$. The wavelength calibration has been done, assuming that the average umbral profile does not exhibit velocities. In the further study we centre on a uniform penumbral part of a large sunspot with radial aligned filaments, presented in Fig.~\ref{Fig1}. 
\vspace{-0.5cm}
   \vspace{-0.5cm}

\section{Inversion procedure}
To derive the physical parameters of the solar atmosphere as a function of continuum optical depth, i.e. temperature $T(\tau$), magnetic field strength $B(\tau)$, field inclination $\gamma(\tau)$, field azimuth $\phi(\tau)$, and line of sight velocity $V_{\rm los}(\tau)$, the SIR code (Ruiz Cobo \& del Toro Iniesta \cite{ruizcobodeltoro92}) has been applied on the spectropolarimetric data set. The values of those parameters are retrieved at a number of optical depth points called nodes. Considering the high-spatial resolution of Hinode observations and assuming that penumbral structures are horizontally resolved, an inversion with only one component has been performed, allowing 5 nodes in $T(\tau)$, 3 nodes in $B(\tau)$ and $V_{\rm los}(\tau)$, and 2 nodes in $\gamma(\tau)$ and $\phi(\tau)$. We have neither considered microturbulent velocities nor stray light contamination. The final synthesised profiles have been convolved with a macroturbulence velocity $V_{\rm mac}$ as an additional free parameter of the inversion.

\section{Determination of a geometrical height scale}
The inversion method delivers for each pixel the stratification of an atmospheric model {\it vs.} continuum optical depth, i.e. we obtain  $\vec{B}(x,y,\tau)$, $T(x,y,\tau)$, etc\dots A geometrical height scale $z(x,y,\tau)$ can be derived by the integration of  
\begin{equation}
d\tau=-\kappa \rho dz \,\,.
\label{optical depth}
\end{equation}
For this integration three ingredients are needed: $\kappa$ (continuum absorption coefficient per gram), $\rho$ (mass density) and the boundary condition  $Z_{W}$=$z$($\tau$=1) (Wilson depression). 

The first ingredient, $\kappa$, is evaluated by SIR from temperatures and abundances. It obviously depends also on gas pressures. However, it can be proven, that this dependence is so small that accurate $\kappa$ values can be obtained from very bad $P_{g}$ values. In fact, SIR obtains the gas pressure stratification by the integration of the hydrostatic equilibrium equation in optical depth scale.

The pressure stratification is used to calculate the second ingredient, the density, using the state equation of an ideal gas and considering partial ionisation. 

If we set $Z_{W}$\,=\,0 at all pixels, a geometric height scale  $z(x,y,\tau)$ can be constructed after integration of Eq.~\ref{optical depth}.  $\vec{B}(x,y,z)$ maps are obtained by interpolation of $\vec{B}(x,y,\tau)$ maps resulting from the SIR inversion. Obviously, such $\vec{B}(x,y,z)$ maps have a divergence different from zero. Furthermore, the models are not in mechanical equilibrium. An optimum choice of $Z_{W}(x,y)$ would minimise at a given height both, the divergence of the magnetic field $\nabla \vec{B}$ and the error in the motion equation.

Neglecting viscosity, the motion equation can be written down as
\begin{equation}
\vec{F}=\vec{J}\times\vec{B}+\rho\,\vec{g}-\nabla P_{g}\,\, .
\label{appmotioneq}
\end{equation}
If we neglect accelerations, $\vec{F}$ must be zero. To ensure the physical meaning of the solution, also $\nabla{\vec{B}}$ must be zero.

We define a merit function as
\begin{equation}
\chi^{2}=\sum_{pixels}{w_{1}(F_{x}^{2}+F_{y}^{2}+F_{z}^{2})+w_{2}(\nabla \vec{B}\nabla \vec{B})}\,\,,
\label{meritfunction}
\end{equation}
where $w_{1}$, $w_{2}$ are coefficients introduced to adequately weight the contribution of both, $\nabla{\vec{B}}$ and $\vec{F}$. By introducing vertical displacements of the atmospheric models at each pixel we try to minimise the merit function (Eq.~\ref{meritfunction}) evaluated at a height level of 200\,km, where the uncertainties of the inverted magnetic field are minimal.
%We want to find the optimun desplacement of the atmospheric models of each pixel which minimize the merit function of Eq \ref{meritfunction}. Therefore we
We use a genetic algorithm changing a chromosome (array $D_{z}(x,y)$ containing the displacements of the atmospheric models at each pixel) until the merit function is minimised. The genetic algorithm has been kindly provided by P\'aez Ma\~n\'a (software engineer at the Instituto de Astrof\'\i sica de Canarias). 
Owing to the uncertainties of the inversion method, the obtained stratification of the magnetic field is neither solenoidal nor satisfying the motion equation. Thus, for the evaluation of $\chi^{2}$ we use slightly modified values $B'(x,y,z)$\,=\,$B(x,y,z)$\,+\,$N_{B}(x,y)$, $\gamma'(x,y,z)$\,=\,$\gamma(x,y,z)$\,+\,$N_{\gamma}(x,y)$, and $\phi'(x,y,z)$\,=\,$\phi(x,y,z)$\,+\,$N_{\phi}(x,y)$, with absolute values of $N_{B}(x,y)$, $ N_{\gamma}(x,y)$, $ N_{\phi}(x,y)$ smaller than the error uncertainties of the respective parameters. The resulting synthesised Stokes profiles, considering $B'$, $\gamma'$ and $\phi'$, are still compatible with the observed Stokes profiles. In summary, the solution found by the genetic algorithm is constituted by the optimum values of $D_{z}(x,y)$, $N_{B}(x,y)$, $N_{\gamma}(x,y)$, and $N_{\phi}(x,y)$.
The best solution is reached setting $w_{1}$and $w_{2}$ such that both addends in Eq. \ref{meritfunction} contribute in an equal manner. In each realisation the code produces slightly different results with a Gaussian distribution around a mean value at each pixel. Therefore, we adopt as final solution the average of 20 individual realisations.

\begin{figure}
 %\sidecaption
   \centering
   \includegraphics[angle=0,width=8cm]{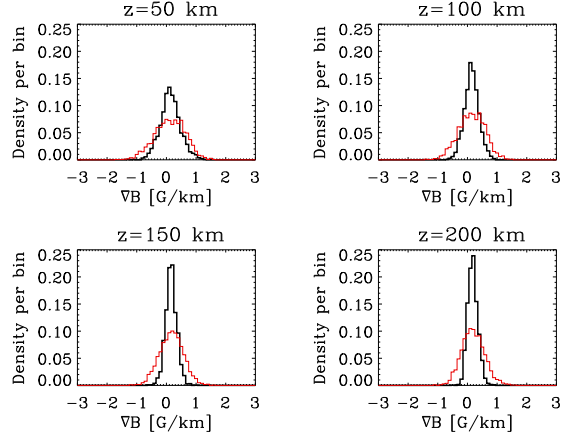}
   \caption{Histograms of the divergence of the magnetic field vector $\nabla \vec{B}$ at four different height layers before (red lines) and after (black lines) the application of the genetic algorithm.  Binsize of 0.1.}
   \label{divcapas}
\end{figure}

After introducing $D_{z}(x,y)$, $N_{B}(x,y)$, $N_{\gamma}(x,y)$ and $N_{\phi}(x,y)$ and interpolating to a common $z$\,-\,scale we can suppose, that the layer at 200\,km approximately satisfies both, $\nabla \vec{B}$ and the motion equation, although the pressure stratification for each pixel continues being the HE one. We can obtain a more accurate $P_{g}$\,-\,stratification by the integration of the $z$\,-\,component of  Eq.~\ref {appmotioneq}. However, when the $P_{g}$\,-\,stratification changes, the $z$\,-\,scale is also modified. Therefore it is easier to integrate this equation in terms of optical depth. Taking into account that both, absorption coefficient $\kappa$ and mean molecular weight $\mu$ hardly depend on gas pressure, we can rewrite the vertical component of  Eq.~\ref {appmotioneq} in terms of $\tau$, setting $\vec{F}$\,=\,0:
\begin{equation}
\kappa\,{\frac{\mu P_{g}}{R T}}\frac{dP_{g}}{d\tau}=g\,\frac{\mu P_{g}}{R T}-{(\vec{J}\times\vec{B})}\mid_{z}\,\, .
\label{motioneqtau}
\end{equation}
After the integration of this equation, e.g. by Runge-Kutta, we obtain $\rho$ and subsequently the new $z$\,-\,scale from  Eq.~\ref{optical depth}. This procedure has to be iterated due to the slight modification of $\vec{B}$ and $\vec{J}$ values. The convergence is reached after just 2 iterations. The models at all pixels are interpolated to a common global $z$\,-\,scale. In Fig.~\ref{divcapas} we present the histograms of $\nabla \vec{B}$ at four different layers. 

\section{Results and Discussion}
\begin{figure}
% \sidecaption
%   \centering
   \includegraphics[angle=0,width=11.5cm]{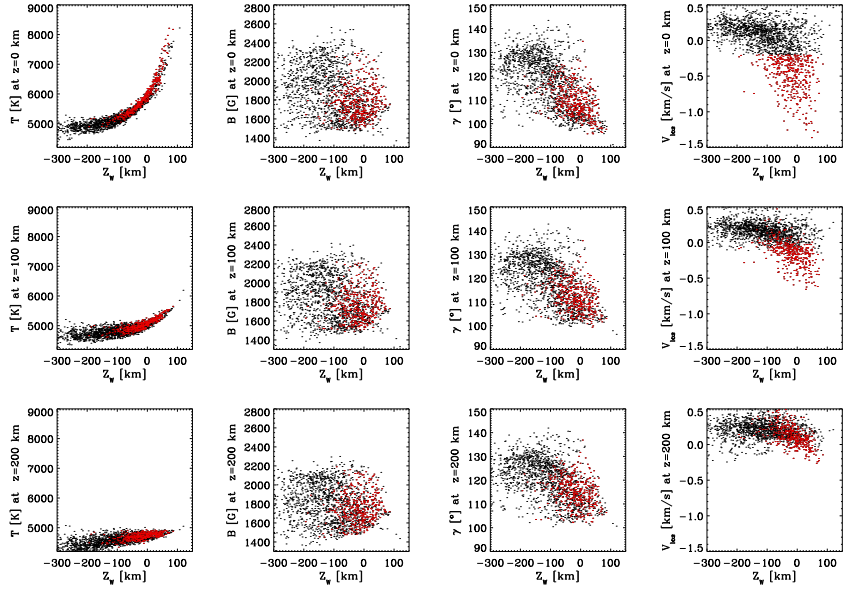}
   \caption{Scatter plots of temperature $T$, magnetic field strength $B$, field inclination $\gamma$, and line of sight velocity $V_{\rm los}$ {\it vs.} Wilson depression; at $z$\,=\,0\,km (upper panels), $z$\,=\,100\,km (middle panels), and $z$\,=\,200\,km (bottom panels). Points with velocities at $z$\,=\,0\,km smaller than -0.2\,km\,s$^{-1}$ are presented in all panels by red colour.}
   \label{scat_zw}
\end{figure}

Using a genetic algorithm we have evaluated the optimum boundary condition for the integration of the vertical component of the motion equation taking into account Lorentz forces. The resulting atmospheric model, interpolated to a common $z$\,-\,scale, shows very small $\nabla \vec{B}$ values in a height range from 50\,km to 200\,km. This demonstrates the consistence of the results, although the genetic algorithm just minimises $\nabla \vec{B}$ at a height of 200\,km. Having at our disposal a common geometrical height scale we can evaluate e.g. electric currents, the Wilson depression, and in general the three dimensional structure of penumbral features. All these topics will be addressed in subsequent papers. Just at a first glance we present in Fig. \ref{scat_zw} scatter plots of several physical quantities at given geometrical heights {\it vs.} the Wilson depression ($Z_{W}$). We observe a strong correlation between temperature $T$ and $Z_{W}$: in places with higher $T$, $\log\tau$\,=\,0 is displaced to upper layers due to the strong dependence of opacity with temperature. Note beside that the $z$\,=\,200\,km layer is nearly isotherm. The magnetic field strength $B$ shows a very weak correlation with $Z_{W}$: higher $Z_{W}$ tends to be related to weaker magnetic fields. Nevertheless, we find clear trends between field inclination $\gamma$ and the Wilson depression, at all layers. The same behaviour is observed in case of line of sight velocity $V_{\rm los}$. In Fig. \ref{scat_zw} we have plotted in red colour points corresponding to pixels harbouring significant values of the vertical component of the Evershed flow ($V_{\rm los}$\,$<$\,-0.2\,km\,s$^{-1}$). Focusing on the distribution of red points throughout the panels we can conclude that the Evershed flow corresponds to areas with increased Wilson depression, hotter temperatures, and weaker and more horizontal magnetic fields. All these properties support the uncombed penumbral model proposed by Solanki \& Montavon \cite{solankimontafon93} (see also Ruiz Cobo \& Bellot Rubio \cite{Ruizcobobellotrubio08}).

%\vspace{-0.5cm}
\begin{acknowledgement}
This work has been supported by the Spanish Ministerio de Educaci\'on y Ciencia through projects ESP 2006-13030-C06-01 and AYA2007-63881.
\end{acknowledgement}
%
%\section*{Appendix}
%\addcontentsline{toc}{section}{Appendix}
%
%
\vspace{-0.5cm}
%%%%%%%%%%%%%%%%%%%%%%%% referenc.tex %%%%%%%%%%%%%%%%%%%%%%%%%%%%%%
% sample references
% %
% Use this file as a template for your own input.
%
%%%%%%%%%%%%%%%%%%%%%%%% Springer-Verlag %%%%%%%%%%%%%%%%%%%%%%%%%%
%
% BibTeX users please use
% \bibliographystyle{}
% \bibliography{}

\begin{thebibliography}{99.}%
% Use the following syntax and markup for your references if 
% the subject of your book is from the field 
% "Mathematics, Physics, Statistics, Computer Science"
%
% Contribution 
  \bibitem{carrolkopf08} Carroll, T. A., \& Kopf, M. 2008, A\&A, 481, L37
  \bibitem{borreroetal05} Borrero, J. M., Lagg, A., Solanki, S. K., \& Collados, M. 2005, A\&A, 436, 333
  \bibitem{borreroetal06} Borrero, J. M., Solanki, S. K., Lagg, A., Socas-Navarro, H., \& Lites, B. 2006, A\&A, 436, 333
  \bibitem{kisugietal07} Kosugi, T., et al. 2007, Sol. Phys. 243, 3
  \bibitem{langhansetal07} Langhans, S., Scharmer, G. B., Kiselman, D., \& L\"ofdahl, M. G. 2007, A\&A, 464, 763
  \bibitem{litesetal01} Lites, B. W., Elmore, D. F., \& Streander, K. V. 2001, ASPC, 236, 33L
  \bibitem{rempeletal08} Rempel, M., Sch\"ussler, M., \& Kn\"olker, M. 2008, arXiv0808.3294
  \bibitem{ruizcobodeltoro92} Ruiz Cobo, B., \& del Toro Iniesta, J. C. 1992, ApJ, 398, 375
  \bibitem{Ruizcobobellotrubio08} Ruiz Cobo, B., \& Bellot Rubio, L. R. 2008, A\&A, 488, 749
  \bibitem{scharmerspruit06} Scharmer, G. B., \& Spruit, H. C. 2006, A\&A, 460, 605
  \bibitem{scharmeretal02} Scharmer, G. B., Gudiksen, B. V.; Kiselman, D., L\"ofdahl, M. G., \& Rouppe van der Voort, L. H. M. 2002, Nature, 420, 151
  \bibitem{solankimontafon93} Solanki, S. K., \& Montavon, C. A. P. 1993, A\&A, 275, 283
  \bibitem{suetterlinetal04} S\"utterlin, P., Bellot Rubio, L. R., \& Schlichenmaier, R. 2004, A\&A, 424, 1049
  \bibitem{vernazzaetal81} Vernazza, J. E., Avrett E. H., \& Loeser, R. 1981 ApJS, 45, 635 (VAL-C)
%
\end{thebibliography}
%

\end{document}